\documentclass[preprint,showpacs,preprintnumbers,amsmath,amssymb, showkeys,superscriptaddress,titlepage]{revtex4-2}
\usepackage{graphicx}
\usepackage{dcolumn}
\usepackage{bm}
\usepackage{hyperref}
\usepackage{textcomp}

\begin{document}
	
	\title{ON FORMATION OF THE $^{12}$C(0$^+_2$) AND $^{12}$C(3$^-$) STATES IN RELATIVISTIC DISSOCIATION OF LIGHT NUCLEI}
	
	\author{\firstname{A.A.}~\surname{Zaitsev}}
	\email{zaicev@jinr.ru}
	\affiliation{Joint Institute for Nuclear Research, Dubna, Russia}%
	\affiliation{P.N. Lebedev Physical Institute of the Russian Academy of Sciences, Moscow, Russia}
	\author{\firstname{P.I.}~\surname{Zarubin}}
	\email{zarubin@jinr.ru}
	\affiliation{Joint Institute for Nuclear Research, Dubna, Russia}%
	\affiliation{P.N. Lebedev Physical Institute of the Russian Academy of Sciences, Moscow, Russia}
	
	
	\begin{abstract}
		The formation of the excited states $^{12}$C(0$^+_2$) and $^{12}$C(3$^-$) is investigated in the dissociation of $^{12}$C $\to$ 3$\alpha$ and $^{16}$O $\to$ 4$\alpha$ at the energy of 3.65 GeV per nucleon in the nuclear emulsion. The identification becomes possible by reconstructing the invariant mass from measurements of emission angles in the approximation of conservation of momentum per nucleon of the parent nucleus. The contribution of the decays $^{12}$C(0$^+_2$) and $^{12}$C(3$^-$) to the dissociation $^{12}$C $\to$ 3$\alpha$ is 11 and 19 \%, and in $^{16}$O $\to$ 4$\alpha$ it is – 20 and 30\%, correspondingly.
	\end{abstract}
	
	\maketitle
	
	\section{Introduction}
	
	Unstable nuclear states with lifetimes several orders of magnitude shorter than the limiting ones for $\beta$-decay (10$^{-12}$ s) and several orders of magnitude longer than the characteristic nuclear scale are weakly bound associations of $\alpha$-particles and nucleons \cite{1}. The decay energies of the above states are noticeably lower than the nearest excitations of parent nuclei with the same nucleon composition. First of all, they include the ground state of the unstable nucleus $^8$Be(0$^+$), decaying into a pair of $\alpha$-particles with the energy of 92 keV and the width of $\Gamma$ = 5.6 eV (or a lifetime of about 10$^{-16}$ s), and then the excitation $^{12}$C(0$^+_2$), known as the Hoyle state, located 285 keV higher than the $^8$Be(0$^+$)$\alpha$ threshold and with the theoretically estimated width of $\Gamma$ = 9.3 eV. The exotic structure of $^8$Be(0$^+$) and $^{12}$C(0$^+_2$), which provides entry into astrophysical nuclear synthesis of $^{12}$C and $^{16}$O, motivates the interest to the mechanism of their formation. The focus of the research is searching for the 4$\alpha$-Bose-Einstein condensate, which is considered to be the excitation of $^{16}$O(0$^+_6$) at 300 keV above the threshold of $^{12}$C(0$^+_2$)$\alpha$ \cite{2}.
	
	The study of the above-mentioned states is the focus of nuclear clustering studies, performed mainly in nuclear reactions at the energy of several MeV per nucleon (reviews \cite{3,4,5}). Focusing on the relativistic fragmentation of nuclei provides a fundamental and visual alternative, since due to the lowest decay energy of $^8$Be(0$^+$) and $^{12}$C(0$^+_2$), they should manifest themselves as pairs and triplets of relativistic $\alpha$-particles with the smallest scattering angles. Despite the undoubted attractiveness of the relativistic approach, only the unique resolution of the nuclear emulsion allows one to carry out such a study \cite{6}.
	
	Being extremely short-lived, unstable states can be identified by the invariant mass $Q$ of the corresponding ensembles of particles. In the general case, $Q$ is determined by the sum of all possible scalar products of the 4-momenta of the decay products $Q$ minus, for convenience, their total mass. This definition is convenient when the number of suitable fragments and particles from the decays coincides. However, the increase in the multiplicity of suitable fragments results in increasing the combinatory of $Q$ when searching for the decays with a smaller number of particles. This difficulty does not appear for $^8$Be and $^{12}$C(0$^+_2$), since the energy of their decays is minimal, and the reflection of more complex excitations at $Q_{3\alpha}$ $<$ 1 MeV is small.
	
	In the fragmentation of nuclei from $^9$Be to $^{197}$Au in the range from hundreds of MeV to tens of GeV per nucleon, the $^8$Be(0$^+$) decays are uniformly identified by determining $Q_{2\alpha}$ from the $\alpha$-particle expansion angles assuming conservation of energy per nucleon of a parent nucleus \cite{6,7}. The current status is summarized in \cite{8}. In the fragmentation $^9$Be $\to$ 2$\alpha$ without combinatorial background, the ground state of $^8$Be(0$^+$) at $Q_{2\alpha}$ $<$ 200 keV and a broad level of $^8$Be(2$^+$) at 3.0 MeV $\Gamma$ = 1.5 MeV for 1 $<$ $Q_{2\alpha}$ $<$ 5 MeV have been identified. These states are present with similar contributions in about three quarters of the events and served as a kind of calibration. The use of the invariant mass has made it possible to identify the 3-body decays of $^9$B in the fragmentation $^{10}$C $\to$ 2$\alpha$2$p$ by $Q_{2\alpha p}$ and $^6$Be in $^7$Be $\to$ $\alpha$2$p$ by $Q_{\alpha2p}$ \cite{9}. In the dissociation $^{12}$C $\to$ 3$\alpha$ and $^{16}$O $\to$ 4$\alpha$, noticeable contributions of $^{12}$C(0$^+_2$) have been detected, and the introduction of the condition on $^8$Be(0$^+$) was not required. The average values of $Q$ corresponded to the established values of the decay energies of $^9$B, $^6$Be and $^{12}$C(0$^+_2$).
	
	The case for the both: the approach itself and its universality in the formation of unstable states, could be strengthened by identifying higher excitations of $^{12}$C \cite{4}, also accompanied by $^8$Be(0$^+$). The first of them is $^{12}$C at 9.64 MeV (+2.37 MeV above the 3$\alpha$ threshold) with $\Gamma$ = 46 keV. Its spin and parity $J^{\pi}$ = 3$^-$ require the transfer of angular momentum to $^{12}$C. The spatial configuration of $^{12}$C(3$^-$) is assumed to be an equilateral triangle of $\alpha$-particles with the unit orbital momentum (in \cite{4}). There is an opportunity of electromagnetic transition: $^{12}$C(3$^-$) $\to$ $^{12}$C(0$^+_2$). The ratio of $^{12}$C(3$^-$) to $^{12}$C(0$^+_2$) may show similarity with the data at low energies, where it is approximately equal to 1.5 (e.g. \cite{3}).
	
	Similarly to $^{16}$O $\to$ $^{12}$C(0$^+_2$)$\alpha$, the channel $^{16}$O $\to$ $^{12}$C(3$^-$)$\alpha$ is possible. However, in this case, at $Q_{3\alpha}$ $>$ 1 MeV, the number of $\alpha$-triples increases sharply. It can be reduced by removing from consideration the events where $^{12}$C(0$^+_2$) or 2$^8$Be(0$^+$) pairs have already been identified. A similar procedure, but with the removal of the events with $^{12}$C(3$^-$) as well, is applicable to the search for the less probable excitations of the $^{12}$C nucleus at 12.7 MeV (+ 5.5 MeV, $\Gamma$ = 18 eV, $J^{\pi}$ = 1$^+$) and 15.1 MeV (+ 7.9 MeV, $\Gamma$ = 44 eV and $J^{\pi}$ = 1$^+$).
	
	The analysis below uses the measurements of the emission angles of relativistic $\alpha$-particles in 510 $^{12}$C $\to$ 3$\alpha$ dissociation events and 641 coherent $^{16}$O $\to$ 4$\alpha$ dissociation events at 3.65 GeV per nucleon (or a momentum of 4.5 GeV/$c$). The measurements were performed in nuclear emulsion exposed at the JINR Synchrophasotron by the emulsion collaboration in the 1970-90s. The statistics were significantly supplemented with respect to $^{12}$C $\to$ 3$\alpha$. The files of these data have been stored in the archive of the BECQUEREL experiment \cite{10}.
	
	\section{Dissociation of $^{12}$C}
	
	In the distribution over $Q_{3\alpha}$ for the dissociation $^{12}$C $\to$ 3$\alpha$, in addition to $^{12}$C(0$^+_2$), some  peculiarity between 2 and 4 MeV has been observed \cite{6}. Inserting the condition $^8$Be(0$^+$) $Q_{2\alpha}$ $<$ 200 keV leads to the $Q_{3\alpha}$ doublet, as shown in Fig. 1 by the dotted line normalized to the number of events \cite{8}. The first peak with the average value $Q_{3\alpha}$(RMS) = 417 $\pm$ 27 (165) keV corresponds to $^{12}$C(0$^+_2$), and the second one, described by the Rayleigh distribution with the parameter $\sigma$ = 2.4 $\pm$ 0.1 MeV – $^{12}$C(3$^-$). For $^{12}$C(3$^-$), the condition 1 MeV $<$ $Q_{3\alpha}$ $<$ 4 MeV has been adopted. First of all, the effect of introducing the condition on $^8$Be(0$^+$) is associated with the suppression of the unidentified contribution of $^{12}$C $\to$ $^8$Be(2$^+$)$\alpha$, and, probably, higher and broader excitations.
	
	\begin{figure}
		\centering
		\includegraphics[width=0.8\textwidth]{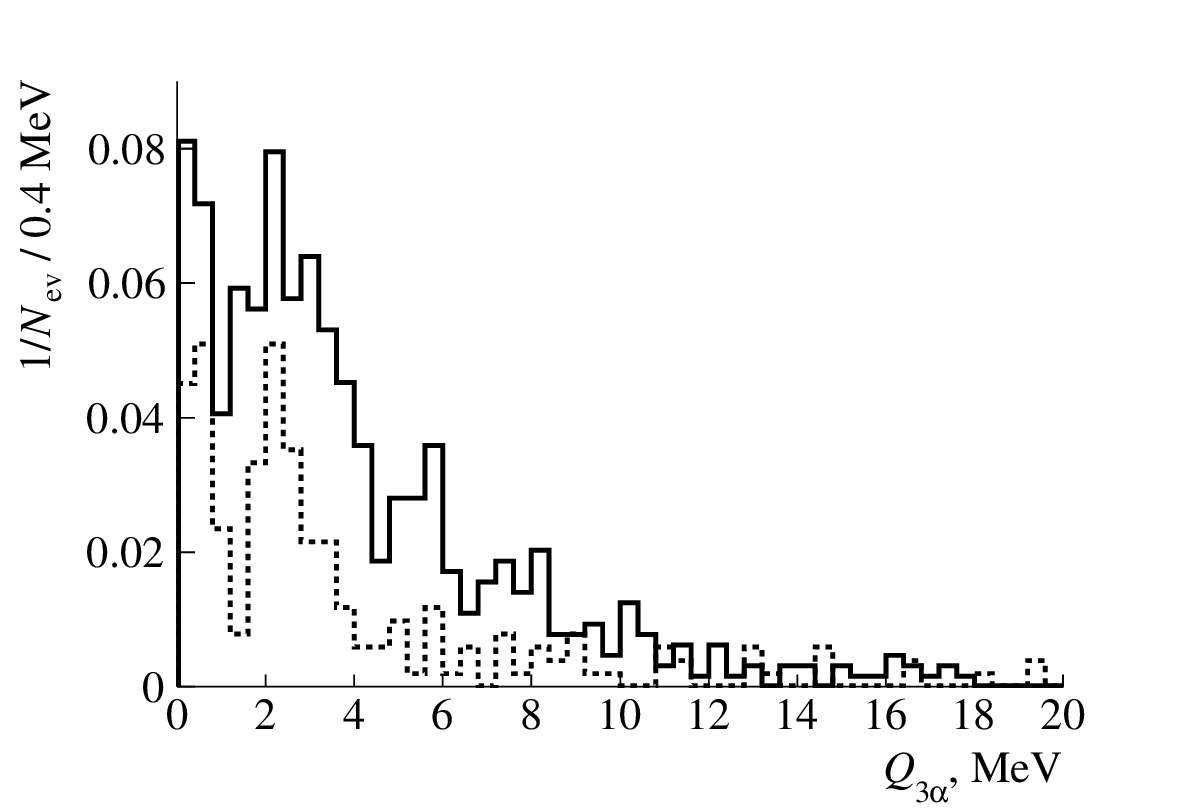}
		\caption{Distribution over $Q_{3\alpha}$ in the events: $^{12}$C $\to$ $^8$Be(0$^+$)$\alpha$ (dots) and $^{16}$O $\to$ $^8$Be(0$^+$)2$\alpha$ (solid); normalized to the number of events.}
	\end{figure}
	
	Then the contributions of $^8$Be(0$^+$), $^{12}$C(0$^+_2$) and $^{12}$C(3$^-$) to the dissociation of $^{12}$C $\to$ 3$\alpha$ are 43 $\pm$ 4, 9 $\pm$ 1, 19 $\pm$ 2\%\%, respectively. The contribution of $^{12}$C(0$^+_2$) decays to the statistics of $^8$Be(0$^+$) is 26 $\pm$ 4\%, and $^{12}$C(3$^-$) – 45 $\pm$ 6\%. The ratio of $^{12}$C(0$^+_2$) and $^{12}$C(3$^-$) is 0.47 $\pm$ 0.06. Making up about a third of the statistics of $^{12}$C $\to$ 3$\alpha$, in the case of $^{12}$C $\to$ $\alpha$$^8$Be(0$^+$), the total contribution of $^{12}$C(0$^+_2$) and $^{12}$C(3$^-$) reaches two thirds.
	
	\section{Dissociation of $^{16}$O}
	
	Despite the growth of combinatory, for the dissociation of $^{16}$O $\to$ 4$\alpha$, there are $^8$Be(0$^+$) and $^{12}$C(0$^+_2$) appeared in the distributions over $Q_{2\alpha}$ and $Q_{3\alpha}$ \cite{6}. When normalized to the number of events, an approximately twofold increase in $^8$Be(0$^+$) and $^{12}$C(0$^+_2$) was found in comparison to the case of $^{12}$C. The main part of the distribution at $Q_{3\alpha}$ $>$ 1 MeV, extending to 20 MeV and having a wide maximum at 1 $<$ $Q_{2\alpha}$ $<$ 5 MeV, is described by the Rayleigh distribution with a parameter of 3.8 MeV. The coincidence of this value (within the errors) with the case of $^{12}$C and the identification of $^{12}$C(3$^-$) in it have indicated the opportunity of the presence of the $^{16}$O $\to$ $^{12}$C(3$^-$)$\alpha$ channel as the leading one.
	
	The condition for the presence of $^8$Be(0$^+$) in the event leads to the signal $^{12}$C(3$^-$) shown in Fig. 1. However, to estimate its contribution, it is desirable to reduce the combinatorial background in $Q_{3\alpha}$. First of all, the already identified events can serve as a source, including 139 $^{12}$C(0$^+_2$)$\alpha$ and 36 2$^8$Be(0$^+$). The distribution of 2$^8$Be(0$^+$) events over $Q_{4\alpha}$ shown in the inset of Fig. 2 is described by the Rayleigh distribution with the parameter 3.6 $\pm$ 0.6 MeV. It results in the distribution of $Q_{3\alpha}$ with the mean value (RMS) of 2.9 $\pm$ 0.13 (1.5) MeV, i.e. just in the very region of $^{12}$C(3$^-$).
	
    \begin{figure}
		\centering
		\includegraphics[width=0.8\textwidth]{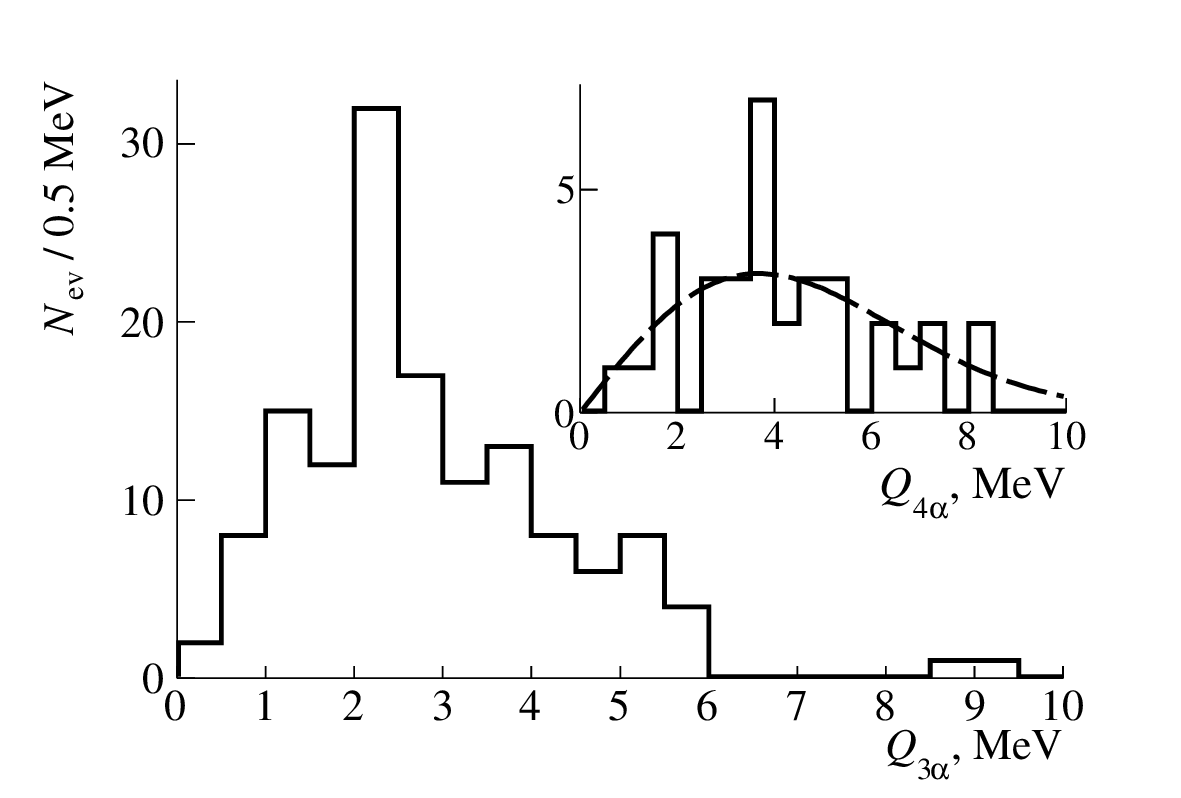}
		\caption{Distribution over $Q_{3\alpha}$ and $Q_{4\alpha}$ (insert) in the events: $^{16}$O $\to$ 2$^8$Be(0$^+$).}
	\end{figure}
	
	Removal of $^{12}$C(0$^+_2$)$\alpha$ and 2$^8$Be(0$^+$) from the analyzed events leads to the $Q_{3\alpha}$ distribution shown in Fig. 3. For comparison, the $Q_{3\alpha}$ distribution is shown for 103 $^{12}$C $\to$ 3$\alpha$ events with a mean value (RMS) of 2.56 $\pm$ 0.08 (0.8) MeV for 1 $<$ $Q_{3\alpha}$ $<$ 4 MeV. There are 196 events in the range accepted for $^{12}$C(3$^-$), with a mean value (RMS) of 2.48 $\pm$ 0.06 (1.0) MeV. Among them, there are 105 having the only ones of single $\alpha$-triplets and 91 - having the double ones, taken into account with a statistical weight of 0.5. The parameters of the $Q_{3\alpha}$ distributions of these groups practically do not differ.
	
	\begin{figure}
		\centering
		\includegraphics[width=0.75\textwidth]{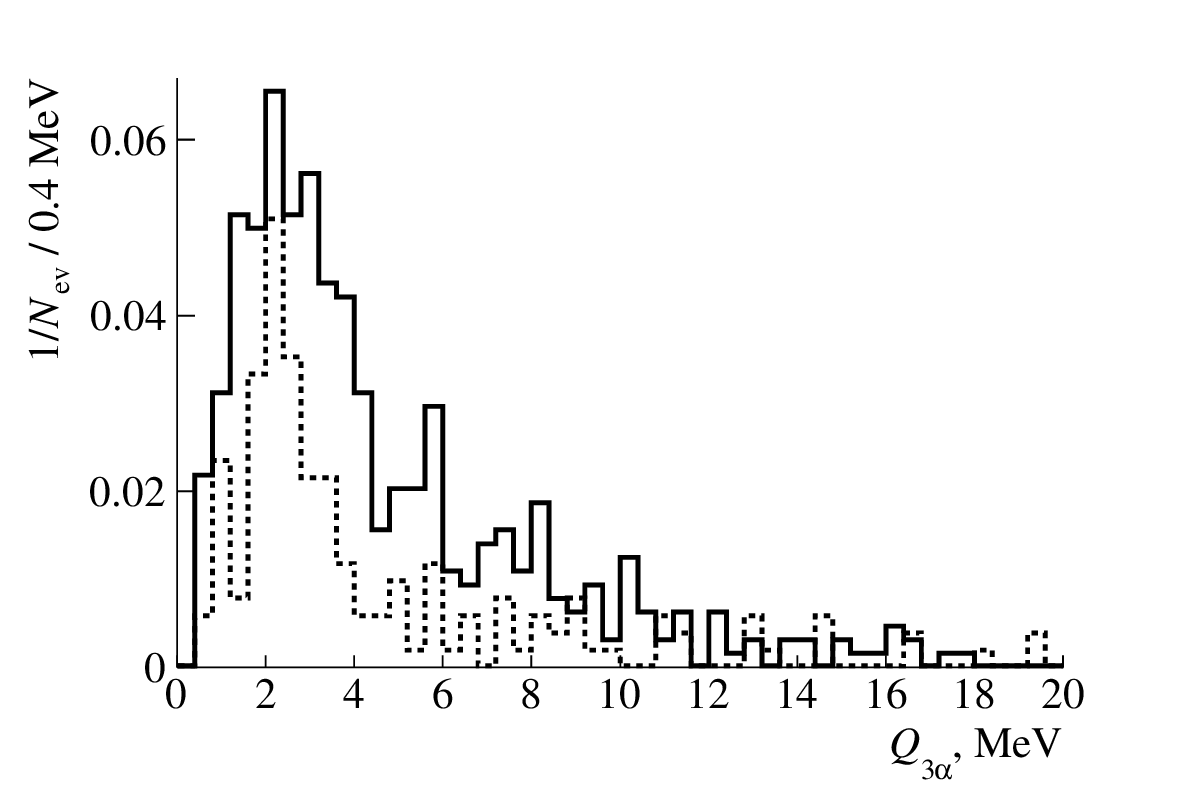}
		\caption{Distribution over $Q_{3\alpha}$ in the events: $^{12}$C $\to$ $^8$Be(0$^+$)$\alpha$ (dots) and $^{16}$O $\to$ $^8$Be(0$^+$)2$\alpha$ (solid) excluding the events with 2$^8$Be(0$^+$) and $^{12}$C(0$^+_2$).}
	\end{figure}
	
	However, the shape of the distribution 1 $<$ $Q_{3\alpha}$ $<$ 4 MeV differs from the case of $^{12}$C serving as a calibration, and the separation of $^{12}$C(0$^+_2$) and $^{12}$C(3$^-$) has deteriorated somewhat. The nearest combinatorial contribution can be given by the events $^{16}$O $\to$ $^8$Be(0$^+$)$^8$Be(2$^+$) which cannot be directly identified. Assuming the equality of $^8$Be(0$^+$) and $^8$Be(2$^+$) in the cases of $^{12}$C and $^{16}$O and their independent formation, the contribution of $^8$Be(0$^+$)$^8$Be(2$^+$) can be estimated to be equal to the contribution of 2$^8$Be(0$^+$). It is subtracted from the number of $^{12}$C(3$^-$) candidates. Taking this remark into account, it can be concluded that $^{12}$C(3$^-$) decays have been identified in the dissociation $^{16}$O $\to$ 4$\alpha$. Then the contribution of the $^{12}$C(0$^+_2$)$\alpha$ channel is 23 $\pm$ 2\%, $^{12}$C(3$^-$)$\alpha$ – 32 $\pm$ 2\%, 2$^8$Be(0$^+$) – 6 $\pm$ 1\%, and the ratio of the $^{12}$C(3$^-$)$\alpha$ and $^{12}$C(0$^+_2$)$\alpha$ channels is equal to: 1.4 $\pm$ 0.1.
	
	\section{Conclusion}
	
	In the dissociation $^{12}$C $\to$ 3$\alpha$ and $^{16}$O $\to$ 4$\alpha$ at 3.65 GeV per nucleon in a nuclear emulsion, the formation of $\alpha$-particle triplets via decays of the excited state of $^{12}$C(3$^-$) has been identified. The identification of $^{12}$C(3$^-$) has become possible by reconstructing the invariant mass from emission angle measurements assuming the conservation of momentum per nucleon of the parent nucleus. Previously, such an approach was successfully used to identify decays of $^{12}$C(0$^+_2$) or the Hoyle state without the condition on the presence of $^8$Be(0$^+$). This condition has improved the $^{12}$C(3$^-$) identification for which the combinatorial background from the $^8$Be(2$^+$)$\alpha$ channel and others has become significant.
	
	The contribution of $^{12}$C(0$^+_2$) and $^{12}$C(3$^-$) decays to the $^{12}$C $\to$ $^8$Be(0$^+$)$\alpha$ channel, which accounts for 43\% of the total statistics, has become 26 and 45\%, respectively. Their ratio is close to the data of low-energy experiments. Thus, $^8$Be(0$^+$) appears mainly as a product of $^{12}$C(0$^+_2$) and $^{12}$C(3$^-$) decays. Although the contribution from higher excitations decaying with $^8$Be(0$^+$) is possible, the statistics do not allow one to make a definitive conclusion. Based on a similar analysis, the $^{12}$C(3$^-$) decays have been identified in the $^{16}$O $\to$ 4$\alpha$ dissociation. The contribution of the $^{12}$C(0$^+_2$)$\alpha$ channel is 22\%, and $^{12}$C(3$^-$)$\alpha$ – 32\%. Thus, the $^{12}$C(3$^-$) respond in the $^{16}$O case is enhanced with respect to the $^{12}$C case, similar to $^{12}$C(0$^+_2$).
	
	These data have indicated the leading role of the $\alpha$-particle excitations of the $^{12}$C and $^{16}$O nuclei in their relativistic dissociation, especially accompanied by $^8$Be(0$^+$). The approach used is optimal to the near-threshold unstable states. There is a prospect of extending the studies of unstable states to neighboring nuclei, including the radioactive ones (e.g., $^{11}$C → $^7$Be$\alpha$). When moving to higher excitations of heavier nuclei, the combinatorial background rapidly increases. The successful identification of $^{12}$C(3$^-$) serves as an argument in favor of observing the 4$\alpha$-particle decays of the $^{16}$O(0$^+_6$) \cite{6}. An alternative, which could be the structure of $^{16}$O(0$^+_6$) as $^{12}$C$\alpha$, is currently being studied. Their overlap cannot be excluded due to the large width of the $^{16}$O(0$^+_6$) level.

\end{document}